\documentclass[preprintnumbers, floatfix,letterpaper,aps,prd,epsfig,nofootinbib,
twocolumn
]{revtex4-1}
\usepackage{bm,graphicx,dcolumn,epstopdf,epsf, latexsym,mathbbol, amssymb,amsmath,color,slashed, mathrsfs,mathcomp, simplewick}
\usepackage[center]{subfigure}
\usepackage{multirow}
\usepackage{makecell}
\usepackage[colorlinks,linkcolor=blue,citecolor=blue,urlcolor=blue]{hyperref}

\begin{document}
\allowdisplaybreaks
 \newcommand{\bq}{\begin{equation}}
 \newcommand{\eq}{\end{equation}}
 \newcommand{\bqn}{\begin{eqnarray}}
 \newcommand{\eqn}{\end{eqnarray}}
 \newcommand{\nb}{\nonumber}
 \newcommand{\lb}{\label}
 \newcommand{\f}{\frac}
 \newcommand{\p}{\partial}
\newcommand{\PRL}{Phys. Rev. Lett.}
\newcommand{\PLB}{Phys. Lett. B}
\newcommand{\PRD}{Phys. Rev. D}
\newcommand{\CQG}{Class. Quantum Grav.}
\newcommand{\JCAP}{J. Cosmol. Astropart. Phys.}
\newcommand{\JHEP}{J. High. Energy. Phys.}
\newcommand{\red}{\textcolor{black}}

\title{Constraining the modified friction in gravitational wave propagation with precessing black hole binaries}

\author{Chunbo Lin}
\affiliation{
Institute for Theoretical Physics \& Cosmology, Zhejiang University of Technology, Hangzhou, 310032, China}
\affiliation{United Center for Gravitational Wave Physics, Zhejiang University of Technology, Hangzhou, 310032, China}

\author{Tao Zhu}
\email{corresponding author: zhut05@zjut.edu.cn}
\affiliation{
Institute for Theoretical Physics \& Cosmology, Zhejiang University of Technology, Hangzhou, 310032, China}
\affiliation{United Center for Gravitational Wave Physics, Zhejiang University of Technology, Hangzhou, 310032, China}

\author{Rui Niu}
\affiliation{Department of Astronomy, University of Science and Technology of China, Hefei, Anhui 230026, China}
\affiliation{School of Astronomy and Space Science, University of Science and Technology of China, Hefei 230026, China}

\author{Wen Zhao}
\affiliation{Department of Astronomy, University of Science and Technology of China, Hefei, Anhui 230026, China}
\affiliation{School of Astronomy and Space Science, University of Science and Technology of China, Hefei 230026, China}

\date{\today}

\begin{abstract}

A broad class of modified gravities can result in a modified friction effect in the propagation of gravitational waves (GWs). This effect changes the amplitude-damping rate of GWs during their propagation in the cosmological distance and thus modifies the standard luminosity distance of GWs in general relativity. Therefore, one can constrain this modified friction by measuring both the luminosity distance and redshift of the GW sources. In this paper, we investigate the prospects of constraining such modified friction effects by using the precessing binary black holes with ground-based GW detectors. For this purpose, we consider 20 precessing events detected by the GW detector network consisting of two LIGO detectors and two third-generation GW detectors (the Einstein Telescope and the Cosmic Explorer). The redshift information of these events is obtained by identifying their possible host galaxies in the GLADE+ galaxy catalog. We show that the precession in the binary system can significantly improve the precision of the luminosity distance and thus lead to a tighter constraint on the modified friction. By assuming narrow priors on cosmological parameters that are consistent with the uncertainties of Planck 2018 results, our analysis shows that the modified friction effect, characterized by two parameters $(\Xi_0, n)$, can be constrained to be $\Xi_0 = 1.002^{+0.004}_{-0.004}$ and $n=3.257^{+2.595}_{-2.192}$, in which the result of $\Xi_0$ is about 2 orders of magnitude better than current results from an analysis with GWTC-3. Our result sets the stage for future research with third-generation GW detectors, offering new insights into gravitational parameter modifications. It also contributes to the understanding of the properties and applications of binary black hole systems with precession.

\end{abstract}

\maketitle

\section{Introduction}
\renewcommand{\theequation}{1.\arabic{equation}} \setcounter{equation}{0}

The field of gravitational wave (GW) astronomy has undergone remarkable growth since the landmark detection of GW150914, a GW event stemming from the merger of two massive black holes, by the LIGO-Virgo Collaboration in 2015 \cite{gw150914}. This groundbreaking discovery opened a new era in gravitational physics. In subsequent years, the LIGO-Virgo-KAGRA (LVK) Collaboration has identified about 90 GW events in the subsequent years \cite{gw-other, gw170817, LIGOScientific:2017ycc, gwtc1, gwtc2, LIGOScientific:2021djp}. These observations are consistent with the gravitational waves generated by the mergers of compact binary systems, as predicted by general relativity (GR). To this end, gravitational waves have proven to be an indispensable tool for testing GR under the extreme conditions of strong gravitational fields and dynamical environments, thereby cementing their significance in the era of GW astronomy \cite{gw150914-testGR, gw170817-testGR, gw170817-speed, testGR_GWTC1, testGR_GWTC2}. In stride with the advancements made by the LVK, the next generation of ground-based GW detectors, including the Einstein Telescope (ET), the Cosmic Explorer (CE) \cite{intro1, intro2}, and space-based ones such as LISA \cite{Robson:2018ifk, LISACosmologyWorkingGroup:2022jok}, Taiji \cite{Ruan:2018tsw, Hu:2017mde}, and Tianqin \cite{Liu:2020eko, Milyukov:2020kyg} are currently in the planning stages. These forthcoming detectors are expected to play a pivotal role in further propelling the frontier of gravitational physics.

While GR stands as the most successful theory of gravity, it encounters substantial difficulties in explaining certain phenomena, including theoretical obstacles like singularities and the challenge of quantizing gravity, as well as observational enigmas such as dark matter and dark energy \cite{in7, in66}. To confront these issues, a comprehensive array of experiments has been carefully crafted to rigorously test GR's theoretical predictions. However, these experiments primarily scrutinize the theory's implications in the weak-field limit \cite{in3, in4, in5}. GWs, on the other hand, arise from regions of strong and highly dynamical gravitational fields and interact weakly with matter, positioning them as powerful tools for probing the nature of gravity \cite{in6}. In light of the GW signals we have detected, a spectrum of alternative gravitational theories has been proposed to address enduring queries within the GR paradigm \cite{MG1, MG2, MG3}. Consequently, the tests of the modified gravities are crucial to ascertaining the definitive theory of gravitational interaction.

A broad class of modified gravity models can introduce a modified friction effect in the propagation of GWs \cite{in17, in18, in19, in8, in9, in10, in11, in12, in13, in14, in15, in16} (see also Table.~\ref{parameters} for a list of modified theories with such an effect). This effect, when it is frequency independent, affects the amplitude-damping rate of GWs during their propagation in the cosmological distance, which in turn changes the standard luminosity distance of GWs in GR. Thus, it is possible to test such an effect with the standard sirens by measuring both the luminosity distance and the redshift of GW sources. The redshift of a GW source can be directly measured if the source has an electromagnetic counterpart. For GW sources without electromagnetic counterparts, the redshift can be inferred using the mass distribution of binary compact objects~\cite{Taylor:2012db, Farr:2019twy, Ezquiaga:2022zkx, MaganaHernandez:2024uty} or by considering potential host galaxies identified in galaxy catalogs~\cite{DES:2020nay, Palmese:2021mjm, Gair:2022zsa}. Constraints on the modified friction effect have been derived through multi-messenger analysis of GW170817, the only binary neutron star merger detected by the LVK Collaboration to date, for which the redshift has been directly measured from the corresponding electromagnetic counterparts \cite{work1, work4, Mastrogiovanni:2020mvm}. Several other constraints have also been obtained by using GW events with redshift information inferred from the host galaxies \cite{Finke:2021aom, Mancarella:2022cgn, Chen:2023wpj}, or the BBH mass function \cite{Mancarella:2021ecn, Leyde:2022orh, Ezquiaga:2021ayr, Mancarella:2022cgn}. Recently, several forecasts for constraining the modified GW friction with future GW detectors have also been carried out in detail (see Refs.~\cite{Liu:2023onj, work10, Chen:2024xkv, Narola:2023viz, Matos:2022uew} and references therein). Note that the extra dimension can also induce modified luminosity distance of GWs, and the corresponding constraints were explored in Refs.~\cite{Pardo:2018ipy, MaganaHernandez:2021zyc}. 

To constrain the modified GW friction, it is essential to measure the luminosity distance and the redshift of the GW source accurately and independently. For a GW source without an electromagnetic counterpart, as we mentioned, an alternative technique is to estimate the source redshift using the statistical redshift information of the possible host galaxies in the galaxy catalogs. To narrow down the number of galaxies in the volume, one needs to improve the measurement precision of both the luminosity distance and the sky location of the source. It has recently been shown in Refs.~\cite{Raymond:2008im, Yun:2023ygz, Green:2020ptm} that the accuracy of luminosity distance estimations can be significantly improved through the observation of GW events from the mergers of precessing binary black holes, and thus can be better candidates for standard sirens \cite{Yun:2023ygz}. The presence of the precession in the binary black hole system has been observed in several GW events in GWTC-3 \cite{Hannam:2021pit, Islam:2023zzj} and can help break the distance-inclination degeneracy to improve the precision of the distance measurement \cite{Raymond:2008im, Yun:2023ygz, Green:2020ptm}. In addition, inclusion of high-order modes (HOMs) in the parameter estimation of GW signals can also help break the distance-inclination degeneracy (see Refs.~\cite{London:2017bcn, Mascioli:2025cnw, Borhanian:2020vyr} and references therein). Moreover, a GW network through the integration of multiple interferometers, both operational and anticipated to be functional (CE and ET), can considerably refine the localization precision of GW events. This enables the simultaneous improvement in the measurement precision of both luminosity distances and redshifts. Consequently, by synergizing the aforementioned precession mechanism with a third-generation ensemble of GW detectors, we stand on the brink of obtaining high-quality luminosity distance and redshift data, which promises to advance the constraints on GW friction beyond the achievements of preceding studies.

In this paper, we investigate the prospects of constraining such modified friction effects by using the precessing binary black holes with a ground-based GW detector network consisting of two LIGO detectors and two third-generation GW detectors (ET and CE). We generate GW signals from merging BBH systems exhibiting precession, utilizing the IMRPhenomPv3 waveform model \cite{4}. These signals are then subjected to a simulated detection process incorporating the combined sensitivities of the two LIGO interferometers, CE and ET. We then determine the redshifts of the host galaxies associated with these GW events within the GLADE+ catalog, leveraging third-generation GW detectors' enhanced spatial localization capabilities. We show that the precession in the binary system can significantly improve the precision of the luminosity distance and thus lead to a tighter constraint on the modified GW friction. 

This paper is organized as follows. We briefly introduce the theory related to modified gravity and the GW friction and give the method of $c_M$ parametrization. In Sec. \ref{s3}, we simulate the GW signal with precession and obtain the luminosity distance from interferometer array measurements. In Sec. \ref{s4}, we illustrate that the third-generation detectors have better luminosity distance and precession estimation accuracies compared to the second-generation detectors by simulating three different detector combinations. Section \ref{s5} is dedicated to the construction of the error box and the determination of the redshift distribution via host galaxies. Sec. \ref{s6} introduces the fundamental statistical framework utilized in this study, encompassing Bayesian analysis and the Markov chain Monte Carlo (MCMC) method, followed by the presentation of GW friction parameter estimation results in Sec. \ref{s7}. The paper concludes with a summary and discussion in Sec. \ref{s8}.

Throughout this paper, the metric convention is chosen as $(-,+,+,+)$, and Greek indices $(\mu,\nu,\cdot\cdot\cdot)$ run over $0,1,2,3$, and Latin indices $(i, \; j,\;k)$ run over $1, 2, 3$. We choose the units $G =c=1$.

\section{Modified GW friction in gravitational propagations}
\renewcommand{\theequation}{2.\arabic{equation}} \setcounter{equation}{0}

In this section, we present a brief review of the modified friction of GWs which can modify the damping rates of the two tensorial modes of GWs. 

\subsection{Modified GW propagations}

We consider the GWs propagating on a homogeneous and isotropic background. The spatial metric in the flat Friedmann-Robertson-Walker universe is written as
\bqn
g_{ij} = a(\tau) (\delta_{ij} + h_{ij}(\tau, x^i)), 
\eqn
where $\tau$ denotes the conformal time, which relates to the cosmic time $t$ by $dt =a d\tau$, and $a$ is the scale factor of the universe. Throughout this paper, we set the present scale factor $a_0 =1$, and $h_{ij}$ denotes the GWs, which represent the transverse and traceless metric perturbations, i.e, 
\bqn
\partial^i h_{ij} =0 = h^i_i.
\eqn
To study the modified frictions in GW propagations, it is convenient to decompose the GWs into circular polarization modes. To study the evolution of $h_{ij}$, we expand it over spatial Fourier harmonics,
 \bqn
 h_{ij}(\tau, x^i) = \sum_{A={\rm R, L}} \int \frac{d^3k}{(2\pi)^3} h_A(\tau, k^i) e^{i k_i x^i} e_{ij}^A(k^i),\nb\\
 \eqn
 where $e_{ij}^A$ denotes the circular polarization tensors and satisfies the relation
 \bqn
 \epsilon^{ijk} n_i e_{kl}^A = i \rho_A e^{j A}_l,
 \eqn
 with $\rho_{\rm R} =1$ and $\rho_{\rm L} = -1$. We find that the propagation equations of these two modes are decoupled, which can be cast into the parametrized form \cite{waveform, Zhu:2023wci}
 \bqn\lb{eom_A}
 h''_A + (2+\bar \nu + \nu_A) \mathcal{H} h'_A + (1+\bar \mu+ \mu_A) k^2 h_A=0,\nb\\
 \eqn
 where a prime denotes the derivative with respect to the conformal time $\tau$ and $\mathcal{H} =a'/a$. 
 
In such a parametrization, the new effects arising from theories beyond GR are fully characterized by four parameters: $\bar \nu$, $\bar \mu$, $\nu_A$, and $\mu_A$. Different parameters correspond to different effects on the propagation of GWs. These effects can be divided into three classes: 1) the frequency-independent effects induced by $\bar \mu$ and $\bar \nu$ which include the modification to the GW speed and friction; 2) the parity-violating effects induced by $\nu_A$ and $\mu_A$ which include the amplitude and velocity birefringence of GWs respectively; and 3) the Lorentz-violating effects induced by frequency-dependent $\bar \nu$ and $\bar \mu$ which include the frequency-independent damping and nonlinear dispersion relation of GWs respectively. The corresponding modified theories with specific forms of the four parameters ${\cal H}\bar \nu$, $\bar \mu$, ${\cal H}\nu_A$, and $\mu_A$ are summarized in Table I in Ref.~\cite{Zhu:2023wci}. The parity- and Lorentz-violating effects, in general, induce amplitude or phase corrections to the GW waveforms of the compact binary inspiral systems since these effects are frequency-dependent \cite{Zhu:2023wci, waveform}. One then can constrain the parity- and Lorentz-violating effects by comparing the modified waveforms with the GW signals (see Refs.~\cite{Zhang:2024rel, Zhu:2023wci, Qiao:2019wsh, Hou:2024xbv, Gong:2023ffb, Zhu:2022uoq, Niu:2022yhr, Gong:2021jgg, Wu:2021ndf, Wang:2020cub, Zhao:2019szi, Li:2024fxy} and references therein). However, these cases are not in the scope of this research, and in this paper, we only focus on the frequency-independent cases.

\begin{table}
\caption{\label{parameters}%
Corresponding parameters ${\cal H} \bar \nu$ and $ \bar \mu$ in specific modified theories of gravity. A comprehensive list of modified gravities with different types of modified GW propagations can be found in Table I of Ref.~\cite{Zhu:2023wci}.}
\begin{ruledtabular}
\begin{tabular}{c|cc}
 Theories of gravity & ${\cal H} \bar \nu$ & $\bar \mu$    \\
  \colrule
  Nonlocal gravity \cite{Belgacem:2017ihm, Belgacem:2018lbp, LISACosmologyWorkingGroup:2019mwx}& $\checkmark$ & ---  \\
Time-dependent Planck mass gravity \cite{Amendola:2017ovw} & $\checkmark$ & ---  \\
Extra dimension (DGP)  \cite{Andriot:2017oaz, Deffayet:2007kf} & \checkmark & ---   \\
$f(R)$ gravity \cite{Hwang:1996xh} & \checkmark & ---  \\
$f(T)$ gravity  \cite{in16} & \checkmark & ---  \\
$f(T, B)$ gravity  \cite{Bahamonde:2021gfp} & \checkmark & ---  \\
$f(Q)$ gravity  \cite{BeltranJimenez:2019tme} & \checkmark & ---  \\
Galileon Cosmology \cite{Chow:2009fm} & \checkmark & ---  \\
Horndeski \cite{horndeski, Bellini:2014fua} &  $\checkmark$ & $\checkmark$ \\
Beyond Horndeski GLPV \cite{Gleyzes:2014qga} & $\checkmark$ & $\checkmark$   \\
DHOST  \cite{Langlois:2017mxy} & $\checkmark$ & $\checkmark$   \\
SME gravity sector  \cite{ONeal-Ault:2020ebv, Nilsson:2022mzq} & $\checkmark$ & $\checkmark$   \\
Generalized scalar-torsion gravity \cite{Gonzalez-Espinoza:2019ajd} & $\checkmark$ & $\checkmark$  \\
Teleparallel Horndeski \cite{Bahamonde:2021gfp} &  --- & $\checkmark$  \\
Generalized TeVeS theory \cite{Sagi:2010ei, Gong:2018cgj} &  --- & $\checkmark$  \\
Effective field theory of inflation \cite{Cheung:2007st} & ---  & $\checkmark$  \\
Scalar-Gauss-Bonnet \cite{Guo:2010jr} & ---  & $\checkmark$  \\
Einstein-\AE{}ether \cite{Oost:2018tcv, Foster:2006az} & ---  & $\checkmark$   \\
Bumblebee gravity \cite{Liang:2022hxd} & ---  & $\checkmark$   \\
Spatial covariant gravities \cite{spatial, Zhu:2022uoq} & $\checkmark$  & $\checkmark$ \\
Gravitational constant variation \cite{Sun:2023bvy} & $\checkmark$  & ---
\end{tabular}
\end{ruledtabular}
\end{table}

\subsection{Modified GW friction and luminosity distance}
  
 When the parameters $\bar \mu$ and $\bar \nu$ are frequency independent, they can induce two distinct and frequency-independent effects on the propagation of GWs. One is the modification to the speed of GWs due to the nonzero of $\bar \mu$, and another effect is the modified friction term of the GWs if $\bar \nu$ is nonzero. These frequency-independent effects can arise from many modified gravities, for example, the scalar-tensor theory, extra dimensions, Einstein-\AE{}ther theory, etc. as summarized in Table \ref{parameters}.

With parameter $\bar \mu$, the speed of the GWs is modified in a frequency-independent manner, $c_{\rm gw} \simeq 1 + \frac{1}{2} \bar \mu$. For a GW event with an electromagnetic counterpart, $c_{\rm gw}$ can be constrained by comparison with the arrival time of the photons. For the binary neutron star merger GW170817 and its associated electromagnetic counterpart GRB170817A \cite{LIGOScientific:2017zic}, the almost coincident observation of both the electromagnetic wave and GW places an exquisite bound on $\bar \mu$, i.e., $-3\times 10^{-15} < \frac{1}{2}\bar \mu <7\times 10^{-16}$. Note that here we set the speed of light as $c=1$. 

The parameter $\bar \nu$ induces an additional friction term on the propagation equation of GWs. In many modified gravities, this term is time-dependent and one can write it in terms of the running of the effective Planck mass in the form of \cite{Lagos:2019kds}
\bqn
{\cal H} \bar \nu = H \frac{d\ln M_*^2}{d\ln a},
\eqn
where $M_*$ is the running of the effective Planck mass. Such a friction term changes the damping rate of the GWs during their propagation. This leads to a GW luminosity distance $d^{\rm gw}_L$ which is related to the standard luminosity distance $d^{\rm em}_L$ of electromagnetic signals as \cite{Belgacem:2018lbp}.
\bqn
d^{\rm gw}_L = d^{\rm em}_L \exp{\Big\{\frac{1}{2}\int_{0}^z \frac{dz'}{1+z'} \bar \nu(z))\Big\}}.
\label{2.7}
\eqn
Thus it is possible to probe such GW friction $\mathcal{H}\bar \nu$ by using the multimessenger measurements of $d_L^{\rm gw}$ and $d_L^{\rm em}$.

However, such a probe relies sensitively on the time evolution of $\mathcal{H}\bar \nu$. There are two approaches to parametrize the time evolution of $\mathcal{H}\bar \nu$: the $c_M$ parametrization \cite{Lagos:2019kds} which is based on the evolution of the dark energy in the Universe and the $\Xi$ parametrization \cite{Belgacem:2018lbp}, which is theory-based parametrization that can fit many modified gravities.

For $c_M$ parametrization, the GW friction is written as \cite{Lagos:2019kds}
\bqn
\bar \nu(z) = c_M \frac{\Omega_\Lambda(z)}{\Omega_\Lambda(0)},
\eqn
where $z$ is the redshift of the GW source and $\Omega_\Lambda$ is the fractional dark energy density. If one considers the dark energy density as a constant, then one has \cite{Leyde:2022orh}
\bqn
\Omega_\Lambda(z) = \frac{\Omega_\Lambda(0)}{\Omega_\Lambda(0)+ (1+z)^3 \Omega_m(0)},
\eqn
where $\Omega_m(0)$ is the value of the fractional energy density of matter. Several constraints on $c_M$ have been derived by using both the information of $d_L^{\rm gw}$ and $d_L^{\rm em}$ from GW events or populations \cite{Leyde:2022orh, Mastrogiovanni:2020mvm, Ezquiaga:2021ayr}. Here we adopt a constraint in Ref. \cite{Leyde:2022orh} from a jointed parameter estimation of the mass distribution, redshift evolution, and GW friction with GWTC-3 for different BBH population models, which gives
\bqn
c_M = -0.6^{+2.2}_{-1.2}.
\eqn
This corresponds to
\bqn
\bar \nu(0)=  -0.6^{+2.2}_{-1.2}.
\eqn

For $\Xi$ parametrization, the full redshift dependence of the GW friction is described by two parameters $(\Xi_0, n)$, with which the ratio between the GW and electromagnetic luminosity distances can be written as \cite{Belgacem:2018lbp}
 \bqn
 \frac{d_L^{\rm gw}(z)}{d_L^{\rm em} (z)} \equiv \Xi(z) = \Xi_0 + \frac{1-\Xi_0}{(1+z)^n}.
 \label{2.12}
 \eqn
Such a parametrization corresponds to
\bqn
\bar \nu(z) = \frac{2 n (1- \Xi_0)}{1-\Xi_0+\Xi_0(1+z)^n}.
\eqn
The relation between the $\Xi$ parametrization and $c_M$ parametrization has been explored in \cite{Mancarella:2021ecn}. Several constraints on $(\Xi_0, n)$ have been obtained using GW events with redshift information inferred from the corresponding electromagnetic counterparts \cite{Mastrogiovanni:2020mvm} or host galaxies \cite{Ezquiaga:2021ayr}, or BBH mass function \cite{Mancarella:2021ecn}. A recent constraint on $(\Xi_0, n)$ was  from an analysis of GW data in GWTC-3 with the BBH mass function, which gives \cite{Mancarella:2021ecn}
 \bqn
 \Xi_0 = 1.0^{+0.6}_{-0.5}, \;\; n=2.5^{+1.7}_{-1.1}
 \label{2.14}
 \eqn
 with a prior uniform in $\ln \Xi_0$. This bound leads to a constraint on $\bar \nu$ in the form of
 \bqn
 -3.0 <\bar \nu(0) < 2.5.
 \eqn

In this paper, we will adopt the $\Xi$ parametrization and explore how future GW observations can improve the current constraints. 

\section{GW generation and luminosity distance measurements}
\label{s3}
\renewcommand{\theequation}{3.\arabic{equation}} \setcounter{equation}{0}

\subsection{GW propagations from BBH merger and detector network}

To date, the LVK Collaboration has reported the observation of about 90 confirmed GW events \cite{1,2}. These events arise from the merging of compact binaries, including binary black holes, binary neutron stars, and black hole-neutron star binaries. A few of these events may possess spin precession \cite{Hannam:2021pit, Islam:2023zzj}. This aspect holds significant potential for advancing research, especially in the realm of GW studies, by refining the accuracy of specific physical parameter determinations \cite{Yun:2023ygz}. Despite the observational challenges associated with black hole precession, recent methodologies, including the introduction of a precession signal-to-noise ratio \cite{4} have demonstrated the feasibility of detecting BBH precessions. In parallel, various theoretical frameworks now integrate precession into their GW simulations. To simulate the GW signals from precessing BBHs, in this paper, we adopt the phenomenological waveform model IMRPhenomPv3 developed in Ref.~\cite{Khan:2018fmp}. This model is built based on its predecessors, such as IMRPhenomC/D, IMRPhenomP, and IMRPhenomPv2 (see Refs.~\cite{Hannam:2013oca, Khan:2015jqa, Husa:2015iqa, Santamaria:2010yb} and references therein) and incorporates a two-spin approach to reflect the latest insights into precession dynamics.

To constrain the modified GW friction using precessing binary black holes with future GW detectors, we focus on the capabilities of a ground-based GW detector network consisting of two LIGO detectors and two third-generation GW detectors (the ET and CE). The specific locations, azimuths of the arms, and lengths of the arms, among other details of the four detectors, are summarized in Table \ref{Tab2}, see also Ref.~\cite{Muttoni:2023prw} for information on the ET and CE. Notably, our configuration utilizes a CE with an arm length of 40 km instead of 20 km, which enables the CE to have a higher detection accuracy. For sensitivity profiles, LIGO is configured with an A+ sensitivity curve \footnote{Our purpose for considering two second-generation detectors alongside next-generation detectors (ET and CE) is to use them for improving the accuracy of GW source localization. As ET and CE begin their observations, it is expected that several second-generation detectors will remain operational. Among these, LIGO A+ stands out as it represents an upgrade positioned between Advanced LIGO and third-generation detectors \cite{Cooper:2022jfr}. Therefore, we included LIGO A+, comprising two detectors, alongside ET and CE in our analysis .}, while CE employs the CE2 (Silicon) sensitivity curve \cite{8}, as forecasted by \textit{gwinc}, and ET utilizes the ET\_D curve \cite{7} for assessing strain sensitivity both in amplitude and spectral density. The analysis is grounded on strain data derived from the background noise and calculated using the power spectral density (PSD) for each interferometer. The sampling rate for these interferometers is set at 2048 Hz, which, according to Nyquist's theorem, establishes the valid signal frequency range as being from 0 Hz to half the sampling rate. Additionally, owing to environmental interferences such as ground vibrations and atmospheric pressure fluctuations, which significantly impede detection below 20 Hz, the minimum frequency threshold for our investigation is established at this value.

Here we would like to mention that our analyses are based on one representative network layout, as mentioned above. We note that while the geometry of the detector network can influence the parameter estimation for individual events, GW sources are expected to be randomly distributed on the sky. As such, when a statistically significant sample of events is considered, the influence of the specific detector configuration is likely to be reduced. The overall combined results should therefore be relatively robust with respect to different network geometries. Exploring the effect of various possible detector configurations in more detail would be an interesting topic for future work.

\begin{table*}[htbp]
	\centering
	\caption{Performance parameters of the GW detector we considered.}
	\label{Tab2}
	\begin{ruledtabular}
		\renewcommand{\arraystretch}{1.4} 
		\begin{tabular}{cccccccc}
			Detector & Abbreviation & Latitude & Longitude & Arm length & $x$-arm azimuth & $y$-arm azimuth & Sensitivity \\
			\hline
		LIGO Hanford & H1 & 46.46$^\circ$ & -119.41$^\circ$ & 4 km & -36.00$^\circ$ & -126.00$^\circ$ & A+ \\
		LIGO Livingston & L1 & 30.56$^\circ$ & -90.77$^\circ$ & 4 km & -107.72$^\circ$ & 162.28$^\circ$ & A+ \\
			Einstein Telescope & ET & 43.70$^\circ$ & 10.42$^\circ$ & 10 km & 139.44$^\circ$ & 79.43$^\circ$ & ET\_D \\
			Cosmic Explorer & CE & -33.29$^\circ$ & 149.09$^\circ$ & 40 km & 135.00$^\circ$ & 45.00$^\circ$ & CE2 Silicon \\
		\end{tabular}
	\end{ruledtabular}
\end{table*}

To quantify the inﬂuence of precession on the distance estimation of binary black holes, we carry out our research using a speciﬁc strategy. As we mentioned, only a few of the GW events in GWTC-3 show possible signals of precession, for example, the events GW200129\_065458 \cite{Hannam:2021pit}, GW190521 \cite{LIGOScientific:2020iuh, LIGOScientific:2020ufj}, and GW191109\_010717 \cite{Zhang:2023fpp}. 
For the third-generation detectors (ET and CE), the detection of precession in binary black holes is expected to be easier. For a rough, conservative estimation using the detection rate of precession in GWTC-3 (about 3 in 90), one expects that at least 500 precessing events could be detected per year within redshift $z \lesssim 0.5$, considering that about $10^4$ GW events can be detected by the third-generation detectors each year for $z \lesssim 0.5$. In this paper, we inject 20 typical precessing binary black hole events for constraining the modified GW friction. These injected datasets are characterized by 15 source parameters $\{m_1, m_2, d_L, \theta_{\rm JN}, {\rm ra}, {\rm dec}, \psi, a_1, a_2, \theta_1, \theta_2, \phi_{\rm JL}, \phi_{12}, t_c, \phi_c\}$ where $m_1$, $m_2$ are the binary black holes' component masses, $d_L$ is the luminosity distance, ${\rm ra}$ and ${\rm dec}$ describe the sky position of the event and $\psi$ is the polarization angle. In addition, $\theta_{\rm JN}$ is the inclination angle of the binary system, and $a_1$ and $a_2$ are the dimensionless spin magnitudes of two black holes. The four angles $\theta_1$, $\theta_2$, $\phi_{\rm JL}$, and $\phi_{12}$ represent the spin misalignment of the binary, which drives the system to precess. Here $t_c$ is the merging time and $\phi_c$ is the coalescence phase. Most of the source parameters of these injected signals are the same as the 20 randomly selected GW events in GWTC-3 with redshifts $z \lesssim 0.5$. The main reason for considering this range is that the redshifts of most galaxies in GLADE+ are less than 0.5. 
The effective precession spin of the precessing binary system is related to the parameters $\theta_1$, $\theta_2$, $a_1$, $a_1$, and $q=m_1/m_2$ as \cite{Hannam:2013oca, Schmidt:2014iyl}
\bqn
\chi_p \equiv {\rm max}\Big\{a_1 \sin \theta_1, \frac{q(4q+3)}{4+3q}a_2 \sin \theta_2\Big\}.
\label{xp}
\eqn
In the 20 injected events, we properly choose the injected values of $\theta_1$, $\theta_2$, $a_1$, and $a_2$ such that their $\chi_p$ is randomly distributed in the range of $[0.3, 0.8]$.
Additionally, the inclination value $\theta_{\rm JN}$ is randomly sampled within the defined domain to ensure the reliability of the simulation results.
In our analysis, similar to the treatment in Ref. \cite{Yun:2023ygz}, we marginalize the parameters $t_c$ and $\phi_c$.

To perform the parameter estimations of the injected GW signals, we adopt the Bayesian parameter estimation tool, \texttt{Bilby} \cite{6}. The priors of the inference parameters of each event, including the masses of the BBHs, the luminosity distance to the GW source, and the orbital inclination of binary systems are assigned default BBH priors in Bilby \cite{6}. We also note that the signal analysis is conducted using the advanced \texttt{dynesty} sampler \cite{Speagle:2019ivv}.

\subsection{Estimations of luminosity distances}
\label{3b}

\begin{figure*}[htbp]
	\centering	\includegraphics[width=1\textwidth]{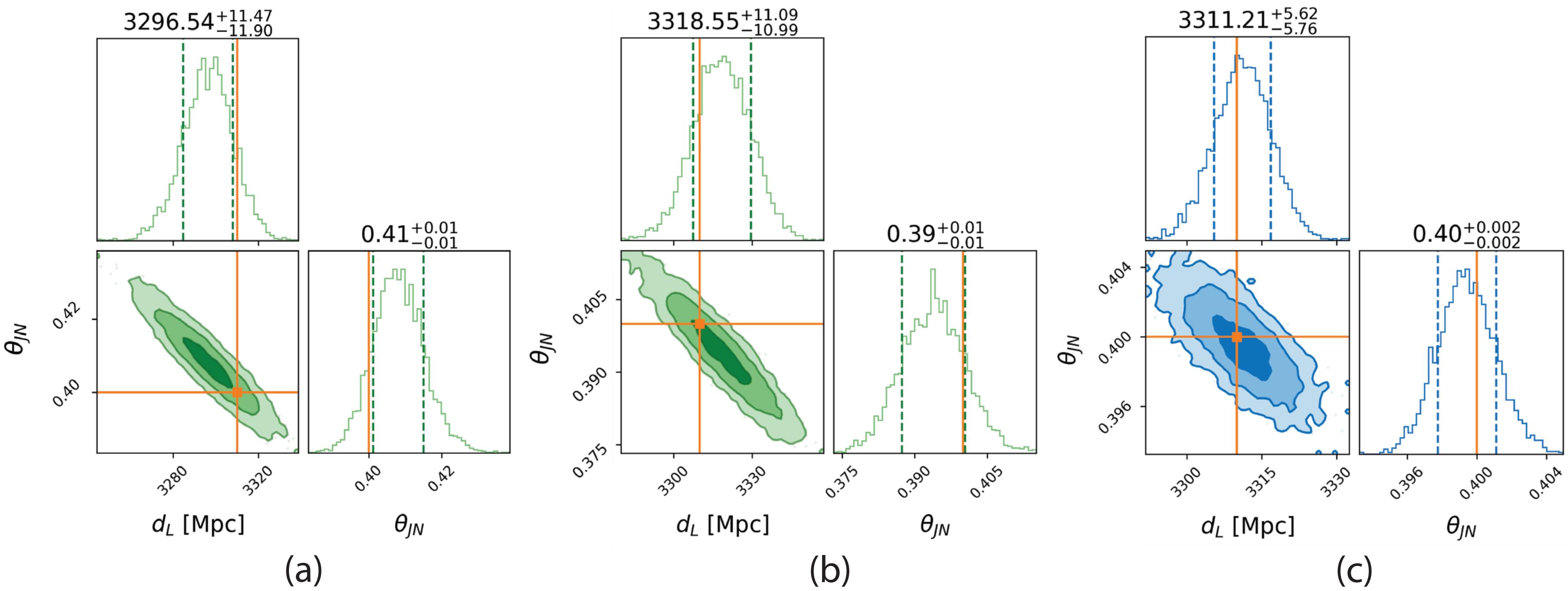}	\caption{Parameter estimation for luminosity distance and the orbital inclination of three injected GW190521-like events: (a) GW190521-like with no precession $\chi_p=0$; (b) GW190521-like with less precession $\chi_p=0.1$; (c) GW190521-like with precession $\chi_p=0.7$.}
	\label{Fig1}
\end{figure*}

We choose to present the results of a typical GW event, the GW190521-like event depicted in Fig.~\ref{Fig1}, which is believed to exhibit obvious precession \cite{LIGOScientific:2020iuh, LIGOScientific:2020ufj}. 
For comparison, we additionally injected two events similar to GW190521, which have a small precession spin $\chi_p =0.1$ and no precession spin $\chi_p =0$, respectively. Figure~\ref{Fig1} consists of three angular plots depicting, in detail, the results of our parameter estimation.
Specifically, Fig.~\ref{Fig1}a-~\ref{Fig1}c elucidate the luminosity distance and orbital inclination parameter estimations derived from the three simulated GW190521-like events with precession spins $\chi_p=0$, $\chi_p=0.1$, and $\chi_p=0.7$, respectively. 
This event is characterized by the merger of black holes with masses of 98.4 and 57.2 solar masses, respectively, as reported in the latest version of this event in the Gravitational Wave Transient Catalog 2 (GWTC-2) \cite{2}. The orange lines in the plots represent the injected values for comparison, while the values above the box plots provide the estimates for these source parameters. 
Notably, Fig.~\ref{Fig1}a (green) illustrates the case with $\chi_p=0$, Fig.~\ref{Fig1}b (green) depicts the case with $\chi_p=0.1$, and Fig.~\ref{Fig1}c (blue) shows the case with $\chi_p=0.7$.
Among the figures, Fig.~\ref{Fig1}a shows the lowest accuracy in luminosity distance estimation, with the injected value (represented by the orange line) falling outside the posterior estimate bounds. The accuracy of Fig.~\ref{Fig1}a is slightly below that of Fig.~\ref{Fig1}b. In contrast, Fig.~\ref{Fig1}c achieves roughly twice the accuracy of the less-precessional case shown in Fig.~\ref{Fig1}b. The introduction of precession contributes to a more accurate estimation of the luminosity distance.
This advancement sets a solid groundwork for subsequent efforts aimed at testing the modified frictions in GW propagations.

The reason for presenting the luminosity distance estimates together with the orbital inclination stems from our objective to illustrate how the degeneracy between these two parameters, evident when the precession effect is not sufficiently large, is mitigated upon incorporating significant precession effects. We observe a notable reduction in the degeneracy between luminosity distance and orbital inclination with the introduction of precession. Similar properties have also been explored and observed in detail in Refs.~\cite{Yun:2023ygz, Green:2020ptm}. This development holds promising prospects for enhancing the analysis of orbital inclination in future investigations.

\section{Comparison of different detector combinations for parameter estimation}
\label{s4}
\renewcommand{\theequation}{4.\arabic{equation}} 
\setcounter{equation}{0}

The main purpose of introducing precession in the analysis is to break the degeneracy between $d_L$ and $\theta_{\rm JN}$, enabling us to obtain a more precise estimate of the luminosity distance. However, the key challenge lies in measuring precession, which is extremely difficult with second-generation detectors due to their relatively lower sensitivity compared to third-generation detectors. As a result, the uncertainties in the estimates of the 15 source parameters mentioned in Sec.~\ref{3b}, obtained through Bayesian analysis, are significantly large. Since precession is calculated based on six of these parameters, as described in Eq.~\eqref{xp}, the uncertainty in precession measurements is consequently also substantial.

\begin{figure*}[htbp]
	\centering
	\includegraphics[width=1\textwidth]{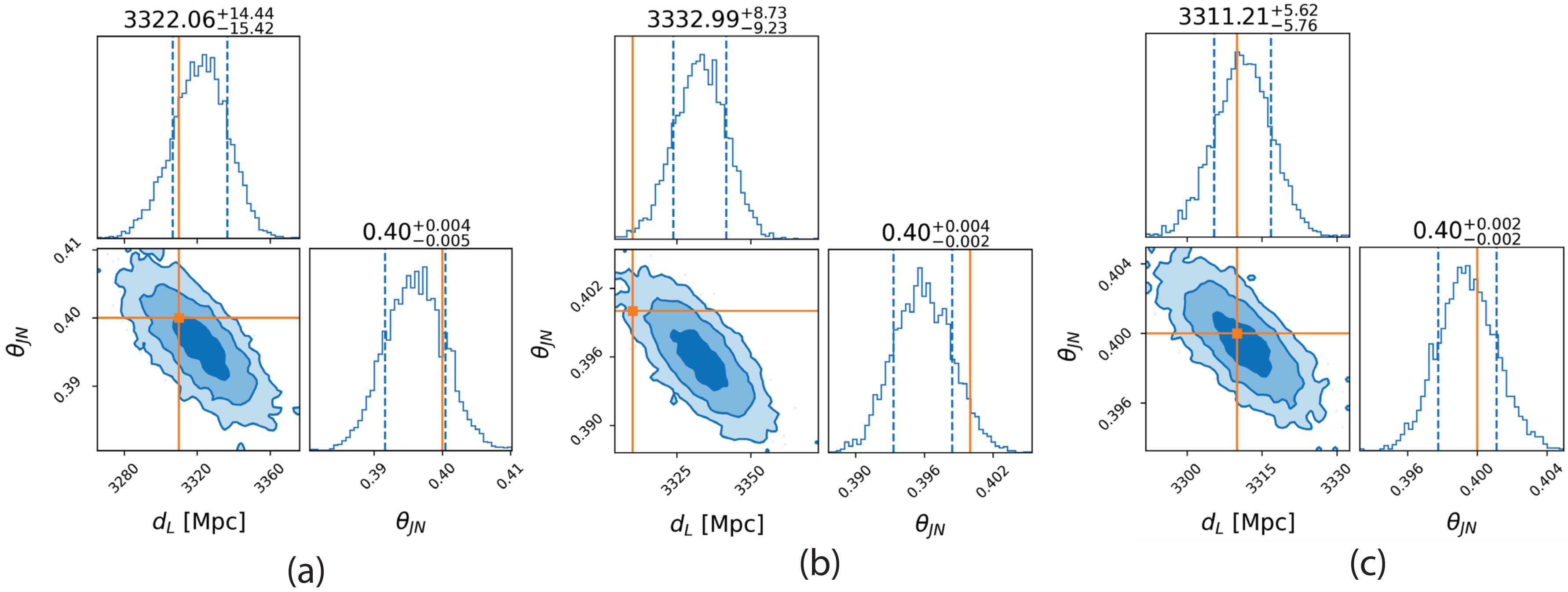}	\caption{Parameter estimation for luminosity distance and the orbital inclination of three injected GW190521-like events with precession spin $\chi_p=0.7$: (a) GW190521-like detected with 2 LIGO detectors; (b) GW190521-like detected with ET and CE; (c) GW190521-like detected with 2 LIGO, ET and CE.}
	\label{Fig3cases}
\end{figure*}

Therefore, it is expected that the high sensitivity of third-generation detectors can help one to accurately measure precession. To illustrate this, we simulate a typical case, a GW150921-like signal observed under different detector configurations. Using this case, we investigate which of the three detector combinations, two LIGOs, ET and CE, or all four detectors, can provide the most accurate measurements of luminosity distance and precession. This is a scenario with an injected value of $\chi_p=0.7$, and Fig.~\ref{Fig3cases} presents the parameter estimation results for $d_L$ and $\theta_{\rm JN}$ in these three configurations.

Fig.~\ref{Fig3cases} consists of three corner plots that provide a detailed overview of our parameter estimation results. Specifically, Fig.~\ref{Fig3cases}a, Fig.~\ref{Fig3cases}b, and Fig.~\ref{Fig3cases}c illustrate the luminosity distance and orbital inclination parameter estimates for three different detector configurations (2 LIGOs, ET and CE, all 4 detectors, respectively) for the same GW event. The orange lines in the plots represent the injected values for comparison, while the values above the box plots provide the estimates for these source parameters.

We can see that the four-detector array configuration yields the best results among the three configurations considered. The 2 LIGO configuration exhibits the largest uncertainty in luminosity distance estimation, approximately three times that of the four-detector array. Although the CE and ET combination shows slightly lower uncertainty, its central estimate deviates significantly from the injected value (3310 Mpc). Consequently, we opted to exclude these two configurations based on a comprehensive evaluation. We also anticipate that the advantages of the four-detector array will become even more pronounced when observing higher redshift targets.

After obtaining the estimation results of 15 source parameters under these three configurations, we calculated the value of $\chi_p$ along with its error bars according to Eq. \eqref{xp} as well. Since Eq. \eqref{xp} is not linear and we have a large number of samples of the 6 parameters used to compute $\chi_p$, it is appropriate to use Monte Carlo Simulation. The precession spin parameter \( \chi_p \) was estimated under three different detector configurations for GW190521-like events. For detection with 2 LIGO detectors, we obtained \( \chi_p = 0.74^{+0.20}_{-0.42} \); for detection with ET and CE detectors, \( \chi_p = 0.67^{+0.06}_{-0.04} \); and for detection with 2 LIGO, ET and CE detectors, \( \chi_p = 0.69^{+0.02}_{-0.03} \). It is evident that the measurement of \( \chi_p \) is significantly more accurate with the four-detector network, achieving a precision that is 12.4 times greater than that of the 2 LIGO configuration and notably higher than that of the ET and CE combination. In the subsequent analysis, we adopt the configuration of the four-detector network.

\section{Redshift distribution of host galaxies}
\label{s5}
\renewcommand{\theequation}{5.\arabic{equation}} \setcounter{equation}{0}

\subsection{The GLADE+ galaxy catalog}

After obtaining the luminosity distances of GW events in the previous section, we acquire independent redshift information utilizing the dark siren method. For the numerous GW sources lacking redshifts ascertainable through optical confirmation, the redshift distribution is typically inferred by identifying the host galaxy \cite{11}. After determining the extent of their spatial orientation through GW observations, we analyze the redshift data from optically observed galaxies within that specific orientation to derive a redshift distribution function. This function serves as the probabilistic distribution for the redshift of the GW source. Thus obtaining catalogs with a high degree of completeness is a prerequisite for searching for host galaxies in localized space. To enhance the realism of this study, instead of generating a simulated galaxy catalog containing information about the redshift distribution of the galaxies, we employed an augmented version of the GLADE galaxy catalog, GLADE+ \cite{12}, as the foundational database for host galaxy identification.
  
\begin{figure}[htbp]
  	\centering
  	\includegraphics[width=0.48\textwidth]{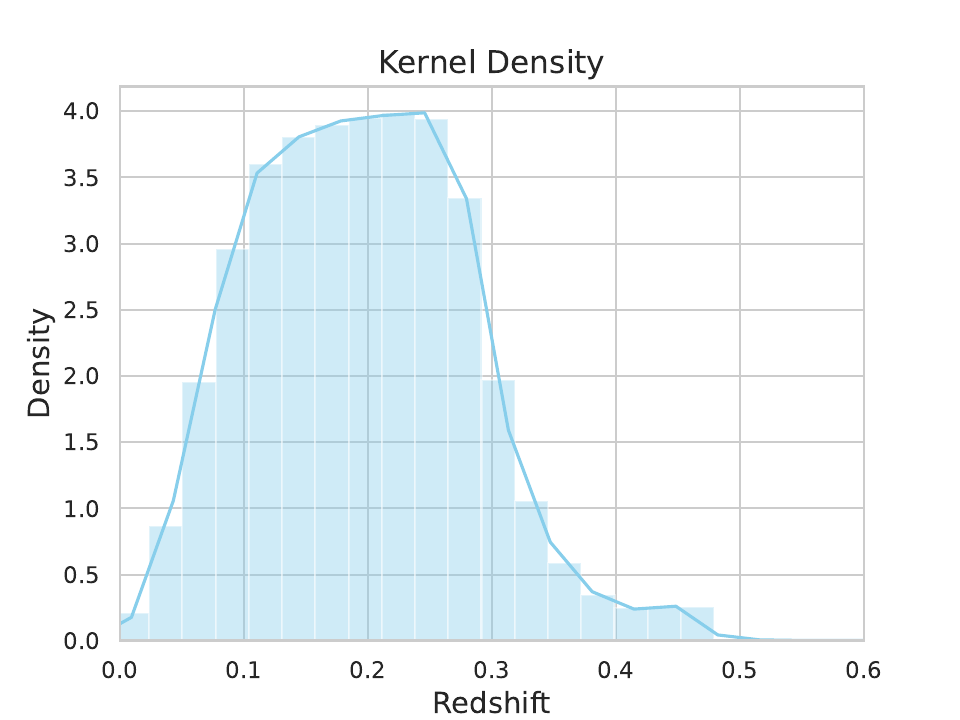}
  	\caption{The kernel density distribution of galaxy redshifts following the preprocessing of the GLADE+ catalog.}
  	\label{Fig2}
  \end{figure}

The GLADE+ catalog encompasses approximately 22.5 million galaxies and around 750,000 quasars \cite{12}. During data preprocessing, galaxies lacking orientation (right ascension and declination) or redshift information were excluded, resulting in a dataset comprising 21,482,830 entries. However, the representation of high-redshift galaxies in GLADE+ is notably incomplete, rendering its completeness inadequate for the requirements of future third-generation GW detectors \cite{13, Yu:2023ico}. Fig.~\ref{Fig2} illustrates the kernel density estimates of the redshifts for all galaxies within the dataset, revealing a significant concentration of GLADE+ galaxies with redshifts spanning from 0.1 to 0.5. To mitigate the impact of catalog data incompleteness on the reliability of the redshift probability distribution for GW sources, we confined our analysis to GW events exhibiting luminosity distances ranging from 900 Mpc to 3000 Mpc. Although this corresponds to a nominal redshift upper limit of $z \lesssim 0.5$, we emphasize that in our final sample, 17 out of the 20 selected events have redshifts below $z = 0.3$, falling well within the redshift range where the GLADE+ catalog maintains reasonable completeness \cite{12}. This selection effectively avoids the regime of severe incompleteness beyond $z \sim 0.3$, as also reflected in the density drop shown in Fig.~\ref{Fig2}. Given the superior precision of redshift data derived from optical observations relative to that of GW information, we assume that the catalog's orientation and redshift for each galaxy are accurate and devoid of errors.
  
  \subsection{Errorbox construction}

From the Bayesian analysis of the injected 20 precessing events using \texttt{Bilby} as described in Sec.~\ref{s3}, we obtained both estimations of the uncertainties of the luminosity distance and sky location for each event. With both the uncertainties, $\Delta \ln(d_{L})$ of luminosity distance and $\Delta \theta^{2}$ of the spatial localization, we define a rectangular error box in the spatial domain where a GW is pinpointed, noting that events at greater luminosity distances correspond to larger error boxes. 

This cuboid error box is conceptually similar to the one used in the statistical dark siren approach of EMRI analysis with LISA data \cite{16}, where redshift information is extracted statistically from a population of galaxies inside a spatial error region. In our case, the box is constructed using the 1$\sigma$ posterior uncertainties of $\ln d_L$ and angular localization (RA, Dec) obtained from full Bayesian parameter estimation using \texttt{Bilby}, instead of analytic scaling formulas.

\begin{figure}[htbp]
  	\centering
  	\includegraphics[width=0.42\textwidth]{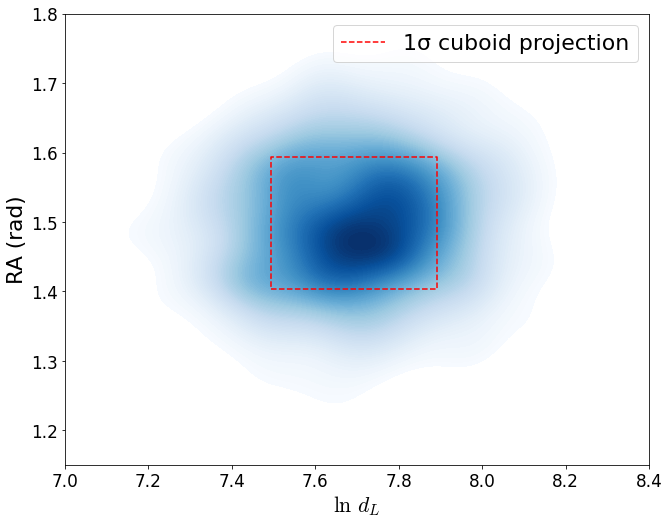}
  	\caption{Posterior distribution of Event A with $1\sigma$ rectangular error box.}
  	\label{fig:posterior_box}
  \end{figure}

We note that our approach implicitly assumes that the posterior distribution in each of these directions is approximately Gaussian, and that the joint distribution factorizes into independent components along the three spatial axes. This approximation enables a computationally efficient statistical cross-matching with galaxy catalogs. Although real posteriors can exhibit skewness or correlations, we find that the cuboid box adequately encloses the high-posterior-density region in most cases. To illustrate our method, we show in Fig.~\ref{fig:posterior_box} the posterior distribution of Event A in the plane of $\ln d_L$ and RA. The red rectangle denotes the $1\sigma$ cuboid error box based on the marginalized standard deviations. We find that although the posterior distribution exhibits mild asymmetry, the box captures the main posterior support region, confirming the practicality of this approach. 

\begin{figure}[htbp]
  	\centering
  	\includegraphics[width=0.42\textwidth]{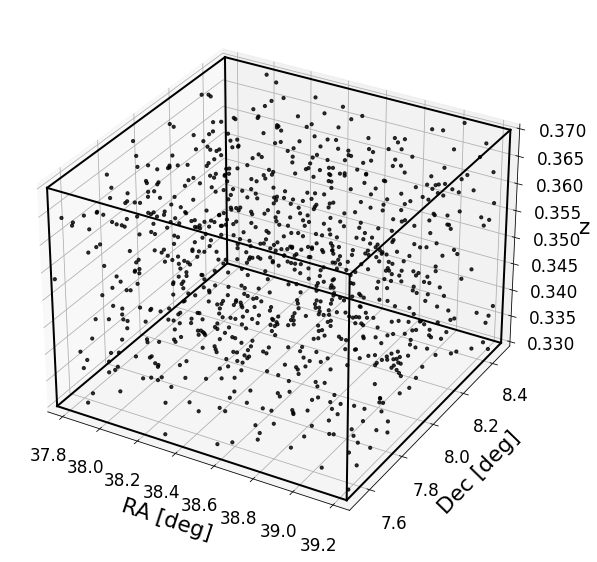}
  	\caption{3D distribution of galaxies in the $1\sigma$ localization volume of a simulated GW event.}
  	\label{fig:posterior_box_3d}
  \end{figure}

In addition, we provide in Fig.~\ref{fig:posterior_box_3d} a three-dimensional visualization of the spatial distribution of galaxies within the cuboid error box for Event B. The axes correspond to RA, Dec, and $\ln d_L$ (or equivalently redshift), and each black dot denotes a galaxy selected from the GLADE+ catalog that falls within the $1\sigma$ bounds of the posterior uncertainties. This representation serves to emphasize the spatial clustering of galaxies within the localization volume. The elongated structure along the line-of-sight direction is evident and reflects the typically larger uncertainty in luminosity distance compared to sky localization. Such plots confirm that even with approximate box-based modeling, the galaxy distribution within the localization volume retains rich information about the redshift structure, which can be exploited in our statistical inference of modified gravity parameters.

Within this framework, we postulate that the redshift of each galaxy within the error box has an equal probability of occurrence \cite{16}. After locating the error box at the GW source, we meticulously traversed all the data in GLADE+ to identify all conceivable galaxies within the designated error box. The numbers of the host galaxies for the 20 events are within a range of 15 to 2000, a testament to the high-precision spatial localization afforded by the collective prowess of the four-detector array.

As shown in Ref.~\cite{Yun:2023ygz}, the luminosity distance estimates for systems exhibiting precession are significantly more precise compared to their non-precessing counterparts. Consequently, within the precession context, the potential number of host galaxies within the error box can be reduced by a factor of 10, in comparison to non-precessing events \cite{Yun:2023ygz}. This reduction directly influences the uncertainty associated with the redshift distribution determination for GW events characterized by precession.

\section{Monte Carlo method (MCMC) for estimation of MODIFIED GW FRICTION}
\label{s6}
\renewcommand{\theequation}{6.\arabic{equation}} \setcounter{equation}{0}

We now independently acquire data on both the redshifts and the luminosity distance distributions of GW events emanating from a sample of 20 BBH mergers. Utilizing the framework provided by the flat ${\rm \Lambda}$CDM cosmological model, we proceed by calculating the standard luminosity distance $d^{\rm em}_L$ of electromagnetic signals
\begin{equation}
d_{L}^{\rm em}=\frac{1+z}{H_{0}} \int_{0}^{z} \frac{d z^{\prime}}{\sqrt{\Omega_\Lambda+\Omega_{m}\left(1+z^{\prime}\right)^3}},
\label{3.1}
\end{equation}
where $H_{0}$ is the Hubble constant, $z$ is the redshift of the GW source and $\Omega_m$ and $\Omega_\Lambda$ are the energy density fractions of matter and cosmological constant, respectively. Note that we assume a flat ${\rm \Lambda}$CDM cosmology in this paper, and thus one has $\Omega_\Lambda + \Omega_m =1$. Combined with equations (\ref{2.7}) and (\ref{2.12}), the luminosity distance of GW with modified friction effect can be expressed as
\bqn
d_{L}^{gw} &=& \left( \Xi_0 + \frac{1-\Xi_0}{(1+z)^n} \right)  \nb\\
&& ~~~~ \times  \frac{1+z}{H_{0}} \int_{0}^{z} \frac{d z^{\prime}}{\sqrt{\Omega_\Lambda+\Omega_{m}\left(1+z^{\prime}\right)^3}}. \nonumber\\
\label{3.2}
\eqn
With the datasets of $d_{L}^{\rm em}$ and $z$ for each event described in the previous sections, we then can start to carry out the analysis of the MCMC implemented by using the open python package, \texttt{emcee}\cite{citeemcee}, to infer the constraints on the parameters $(\Xi_0, n)$. For this purpose, we explore four parameters, $(H_0, \Omega_m, \Xi_0, n)$ to fit the luminosity distances of GW described in (\ref{3.2}) to the datasets of $d_{L}^{\rm em}$ and $z$. 

To carry out our MCMC analysis with the above parameter space, we use uniform prior distributions for all four parameters $(H_0, \Omega_m, \Xi_0, n)$. Specifically, we consider two types of prior ranges, the wide prior and the narrow prior for the cosmological parameters $(H_0, \Omega_m)$, respectively. In the wide prior, we use a wide range of cosmological parameters $(H_0, \Omega_m)$, while in the narrow prior the range of $(H_0, \Omega_m)$ is set to be consistent with the uncertainties of Planck 2018 results \cite{20}. The prior sets used for our MCMC analysis are summarized in Table.~\ref{priors}. In addition, we construct the likelihood function ${\cal L}$ based on (\ref{3.1}) and (\ref{3.2}) for our analysis as follows,
\bqn
\ln \mathcal{L}=-\frac{1}{2} \sum_{i} \frac{\left(d_{L_{\text{obs}, i}}-d_{L_{\text{theo}, i}}\right)^{2}}{\sigma_{i}^{2}}.
\eqn
The observed luminosity distance for the $i$th data point, symbolized as $d_{L_{\text{obs},i}}$, is derived from the analysis of received GW signals. The corresponding theoretical luminosity distance, computed utilizing the model parameters ($H_0$, $\Omega_m$, $\Xi_0$, $n$) in conjunction with the redshift $z_i$, which is ascertained from the host galaxies of the GW events as expounded previously, is expressed as $d_{L_{\text{theo},i}}$, refer to formula (\ref{3.2}). The total squared uncertainty for the $i$th observation, incorporating both the redshift error $xerr_i$ and the luminosity distance error $yerr_i$, is quantified by $\sigma_{i}^2 = xerr_i^2 + yerr_i^2$. Summation across the entirety of observed data points is denoted by $\sum_i$.

\begin{table}
\caption{\label{priors}%
Two types of uniform prior sets were used for our MCMC analysis.}
\begin{ruledtabular}
\renewcommand{\arraystretch}{1.2}
\begin{tabular}{ccc}
 Parameters & wide prior & narrow prior    \\
  \colrule
$H_0$ (${\rm Mpc}^{-1}$) & [60, 90] & [66.9, 67.9]  \\
$\Omega_m$  & [0.1, 0.5] & [0.308, 0.322]  \\
$\Xi_0$  & [0.5, 1.5] & [0.5, 1.5]  \\
$n$ & [0.5, 7] & [0.5, 7]  \\
\end{tabular}
\end{ruledtabular}
\end{table}

Consequently, we can deduce the posterior probability distribution for the cosmological parameters \(P(H_0, \Omega_m, \Xi_0, n)\) from our observational datasets by utilizing the Bayesian inference. The posterior probability is described by the following equation,
\bqn
&& P(H_0, \Omega_m, \Xi_0, n | \text{data}) \nb\\
&&~~~~~ \propto \mathcal{L}(\text{data} | H_0, \Omega_m, \Xi_0, n)  P(H_0, \Omega_m, \Xi_0, n).\nb\\
\eqn
Herein, \(P(H_0, \Omega_m, \Xi_0, n | \text{Data})\) denotes the posterior probability distribution of the parameters given the data, \(\mathcal{L} (\text{Data} | H_0, \Omega_m, \Xi_0, n)\) signifies the likelihood of the data under the specified parameters, and \(P(H_0, \Omega_m, \Xi_0, n)\) represents the prior knowledge of the parameters.

\section{RESULTS AND DISCUSSION}
\label{s7}
\renewcommand{\theequation}{7.\arabic{equation}} \setcounter{equation}{0}

\begin{figure}[htbp]
	\centering
	\includegraphics[width=0.48\textwidth]{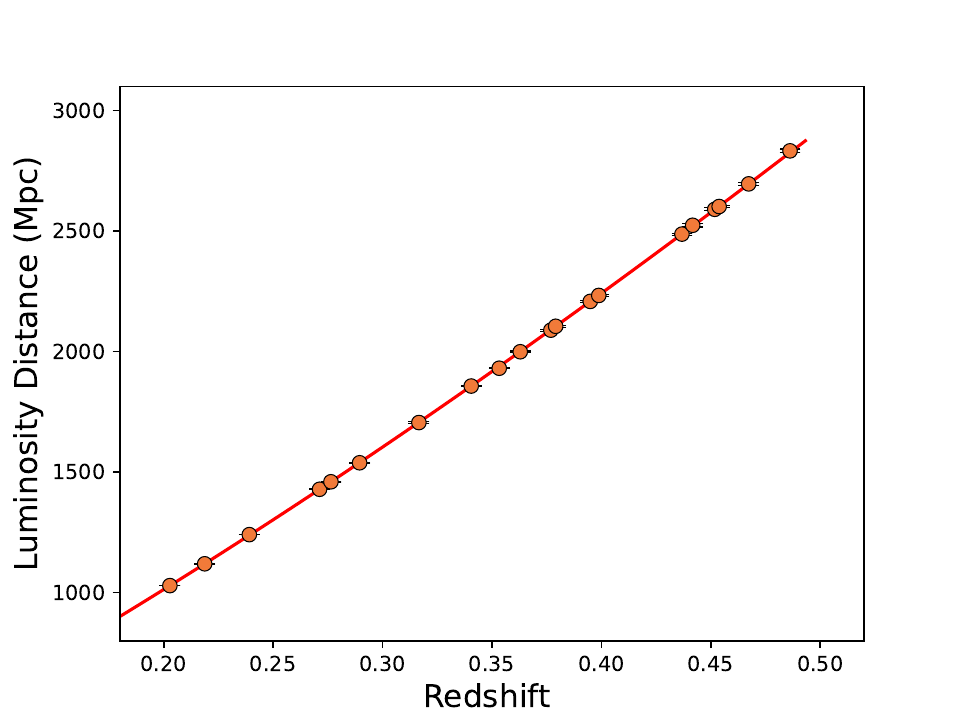}
	\caption{Scatter plot of the luminosity distance-redshift distribution for 20 BBH merger GW events, with error bars.}
	\label{Fig3}
  \end{figure}

Upon deriving luminosity distances from the GW signals received, and independently determining their redshift distributions through the identification of host galaxies, we depicted the data for 20 GW events on a scatter plot in Fig.~\ref{Fig3}.  The horizontal axis of Fig.~\ref{Fig3} represents the redshifts of GWs, whereas the vertical axis denotes their luminosity distances. Furthermore, we applied a cosmological model that incorporates the modified GW friction term to these data, which is represented by a red curve in Fig.~\ref{Fig3}. Owing to the constraints of the GLADE+ catalog, the redshifts for our GW event dataset predominantly range between 0.2 and 0.5. Because the redshift data were obtained by the dark sirens method, the redshift is essentially a statistical distribution. The values of the 50th quantile of this distribution are illustrated in Fig.~\ref{Fig3}.

We then incorporate data of both the redshifts and luminosity distances into eq.~(\ref{3.2}) via the MCMC method to impose constraints on ($H_0$, $\Omega_m$, $\Xi_0$, $n$). We perform the MCMC analysis with two types of prior sets in Table.~\ref{priors} separately, and thus in what follows we present their results in different subsections.

\subsection{Results with wide prior}

\begin{figure}[htbp]
  	\centering
  	\includegraphics[width=0.45\textwidth]{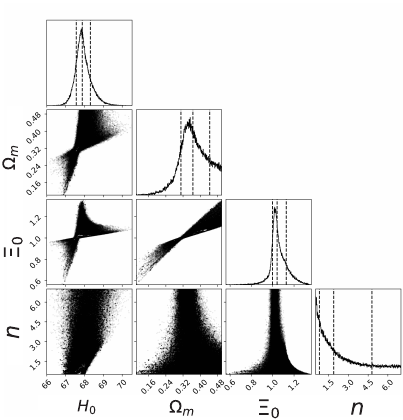}
  	\caption{Corner plot illustrating the constraints on the four parameters $(H_0, \Omega_m, \Xi_0, n)$ derived from GW data with wide prior on $(H_0, \Omega_m)$. The plot features a black line indicating the uncertainties based on the 16th, 50th, and 84th percentiles of the samples in the marginalized distributions.}
  	\label{Fig4}
\end{figure}

Let us first present the results with wide priors. We explore the parameter space of $(H_0, \Omega_m,\Xi_0, n)$ through the MCMC analyses with 20 precessing GW events. The full posterior distributions of the four parameters $(H_0, \Omega_m,\Xi_0, n)$ from the three analyses are depicted in the corner plots of Fig.~\ref{Fig4}. We also present the 68\% C.L bounds on the four parameters for each analysis in Table.~\ref{wide}. The MCMC analysis leads to
 \bqn
  \Xi_0 = 1.041^{+0.086}_{-0.043}, \;\; n=1.879^{+2.909}_{-1.091}.
  \eqn 
This bound leads to a constraint on $\bar \nu(0)$,
  \bqn
-0.368 < {\bar \nu}(0) < 0.008.
  \eqn
We observe that the precision on $\Xi_0$ is improved by a factor of 8.5, compared to that from the analysis with GWTC-3.
  
\begin{table}
	\caption{Estimation results of parameters under different priors.}
	\label{wide}
	\begin{ruledtabular}
		\renewcommand{\arraystretch}{1.3} 
		\begin{tabular}{cccc}
			& wide prior & narrow prior  \\
			\colrule
			$H_0$ (${\rm Mpc}^{-1}$) & $67.881^{+0.430}_{-0.319}$ & $-$  \\
			$\Omega_{\Lambda}$ & $0.366^{+0.078}_{-0.056}$ & $-$  \\
			$\Xi_{0}$ & $1.041^{+0.086}_{-0.043}$ & $1.002^{+0.004}_{-0.004}$  \\
			$n$ & $1.879^{+2.909}_{-1.091}$ & $3.257^{+2.595}_{-2.192}$  \\
		\end{tabular}
	\end{ruledtabular}
\end{table}
   
\subsection{Results with narrow prior}

\begin{figure}[htbp]
 	\centering
 	\includegraphics[width=0.47\textwidth]{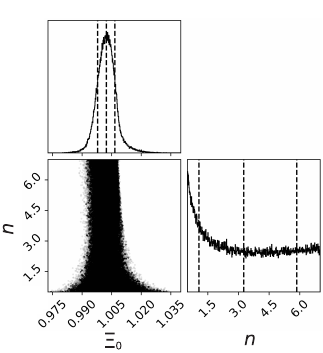}
 	\caption{This corner plot delineates the constraints on the two parameters: $\Xi_0$ and $n$, under narrow prior conditions for the parameters $H_0$ and $\Omega_m$. The uncertainties, derived from the 16th, 50th, and 84th percentiles of the samples in the marginalized distributions, are marked with a black line akin to Figure \ref{Fig4}.}
 	\label{Fig5R}
   \end{figure}

\begin{figure*}[htbp]
  	\centering
  	\includegraphics[width=0.90\textwidth]{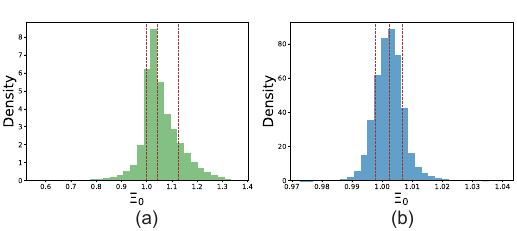}
  	\caption{In Figs.~\ref{Fig6R}a (rendered in green), we present the constraints derived from an analysis of 20 GW events under wide prior conditions. Conversely, Fig.~\ref{Fig6R}b (illustrated in blue) delineates the constraints analyzed under narrow prior conditions for the parameters $H_0$ and $\Omega_m$. }
  	\label{Fig6R}
  \end{figure*}

We then turn to present the results with the narrow prior. In this case, we explore the parameter space of $(H_0, \Omega_m,\Xi_0, n)$ through the MCMC analyses with 20 injected precessing events with priors on $(H_0, \Omega_m)$ that are consistent with the uncertainties of Planck 2018 results \cite{20}. The posterior distributions of the parameters $(\Xi_0, n)$ from the analysis are depicted in the corner plots of Fig.~\ref{Fig5R}. The 68\% C.L bounds on the two parameters $(\Xi_0, n)$ for the analysis are presented in Table.~\ref{wide}. The analysis gives
\bqn
\Xi_0 = 1.002^{+0.004}_{-0.004}, \;\; n=3.257^{+2.595}_{-2.192}.
  \eqn 
One observes that this bound improves that with wide prior by a factor of 18, and improves the current constraint from an analysis with GWTC-3 about two orders of magnitude better. The above bound leads to a constraint on $\bar \nu(0)$,
\bqn
-0.041 < {\bar \nu}(0) < 0.011.
\eqn
In addition, we also plot Fig.~\ref{Fig6R} to compare the marginalized posterior distribution of the parameter $\Xi_0$ from the analysis with wide prior and narrow prior, respectively.

In summary, we show that, with the ET and CE operational, the observation of only 20 GW signals from precessing BBH mergers could refine the current precision of the $\Xi_0$ constraint by a minimum of 8.5 times for wide prior and by at least two orders of magnitude for the narrow prior.

\section{conclusion}
\label{s8}
 \renewcommand{\theequation}{8.\arabic{equation}} \setcounter{equation}{0}
 
In this paper, we delve into the investigation of the modified friction effect in the GW propagations. This effect modifies the amplitude-damping rate of GWs and can arise from a broad class of modified gravities. Due to the modified damping rate of GWs, the corresponding luminosity distance of the GW source is different from that in GR. Therefore, one can test this effect with the GW standard sirens, by measuring both the luminosity distance and the redshift of GW sources. Our purpose here is to evaluate the capability of the next generation of ground-based
(aLIGO, CE, and ET) detectors in constraining this effect with mergers of precessing binary black holes. 

We begin by injecting 20 precessing GW events detected by the GW detector network consisting of two LIGO detectors and two third-generation GW detectors (ET and CE). Through a Bayesian analysis of these injected signals by using the open source package \texttt{bilby}, we estimate both the luminosity distances and the uncertainties of the sky locations of these events. With these results, we then independently ascertained their redshift distributions by identifying host galaxies via the preprocessed GLADE+ catalog. Utilizing the inferred redshifts and luminosity distances of the 20 precessing events, we perform the MCMC analysis to place constraints on the modified GW friction.

Specifically, we consider two types of priors on the cosmological parameters $(H_0, \Omega_m)$ in the MCMC analysis, the wide prior and the narrow prior. In the wide prior, we use a wide range of cosmological parameters $(H_0, \Omega_m)$, while in the narrow prior the range of $(H_0, \Omega_m)$ is set to be consistent with the uncertainties of Planck 2018 results \cite{20}. Our analyses show that the constraint on the modified friction parameter $\Xi_0$ from wide prior is about 8.5 times better than the current result from the analysis with GWTC-3, while the result from narrow prior is about two orders of magnitude better.

This investigation offers a robust estimation of the constraints on modified GW friction in the context of third-generation detectors, establishing a groundwork for subsequent research following the operational commencement of these detectors. Furthermore, our findings may inspire novel approaches for imposing constraints on the parameters of modified gravity theories, facilitating further research into the unique characteristics and effects of precessing BBH merger systems.

Current study only focuses on waveforms incorporating precession effects to investigate their impact on parameter estimation.  Here we would like to mention that an important direction for future research is to extend our analysis by simultaneously including HOMs and precession effects. Detectable HOMs are known to effectively break distance-inclination degeneracies \cite{London:2017bcn, Mascioli:2025cnw, Borhanian:2020vyr}, which can significantly improve the accuracy of gravitational wave luminosity distance measurements. By combining both precession and HOMs, it may be possible to achieve even tighter constraints on modified GW friction, thereby enhancing our ability to test deviations from general relativity. We leave this promising prospect for future investigation.

\section*{Acknowledgments}

This work is supported in part by the National Key Research and Development Program of China Grant No. 2020YFC2201503, the Zhejiang Provincial Natural Science Foundation of China under Grant No. LR21A050001 and No. LY20A050002, the National Natural Science Foundation of China under Grant No. 12275238, and the Fundamental Research Funds for the Provincial Universities of Zhejiang in China under Grant No. RF-A2019015. WZ is supported by the National Key R\&D Program of China Grant No. 2021YFC2203102 and 2022YFC2204602, Strategic Priority Research Program of the Chinese Academy of Science Grant No. XDB0550300, NSFC No. 12325301 and 12273035, the Fundamental Research Funds for the Central Universities under Grant No. WK3440000004.

\appendix

\end{document}